\documentclass[aps,prd,twocolumn,superscriptaddress,nofootinbib,preprintnumbers]{revtex4-1}
\usepackage{mathrsfs}
\usepackage{amsfonts}
\usepackage{amsmath}
\usepackage{array}
\usepackage{verbatim}
\usepackage{epsfig}
\usepackage{graphicx}
\usepackage{color}

\newcommand{\beq}{\begin {equation}}
\newcommand{\eeq}{\end   {equation}}
\newcommand{\nn }{\nonumber        }

\begin{document}

\preprint{
{\vbox {
\hbox{\bf MSUHEP-18-018}
}}}
\vspace*{0.2cm}

\title{Resummation of High Order Corrections in Z Boson Plus Jet Production at the LHC}

\author{Peng Sun}
\email{pengsun@msu.edu}
\affiliation{Department of Physics and Institute of Theoretical Physics, Nanjing Normal University, Nanjing, Jiangsu, 210023, China}

\author{Bin Yan}
\email{yanbin1@msu.edu}
\affiliation{Department of Physics and Astronomy, Michigan State University,
East Lansing, MI 48824, USA}

\author{C.-P. Yuan}
\email{yuan@pa.msu.edu}
\affiliation{Department of Physics and Astronomy, Michigan State University,
East Lansing, MI 48824, USA}

\author{Feng Yuan}
\email{fyuan@lbl.gov}
\affiliation{Nuclear Science Division, Lawrence Berkeley National
Laboratory, Berkeley, CA 94720, USA}

\begin{abstract}
We study the multiple soft gluon radiation effects in $Z$ boson plus
jet production at the LHC. By applying the transverse momentum
dependent  factorization formalism, the large logarithms introduced by the small
total transverse momentum of the $Z$ boson plus leading jet final state system,
are resummed
to all orders in the expansion of the strong interaction coupling at the accuracy of Next-to-Leading Logarithm(NLL).
We also compare the prediction of our resummation calculation to the CMS data by employing a reweighting procedure to estimate the effect from imposing kinematic cuts on the leptons from $Z$ boson decay, and find good agreement for both the  imbalance transverse momentum and the azimuthal angle correlation of the final state $Z$ boson and  leading jet system, for $pp\to Z+jet$ production at the LHC.  
\end{abstract}

\maketitle
\noindent{\bf Introduction.}
The $Z$ boson and jets associated production  at Large Hadron Collider (LHC)
plays an important role in our knowledge of the Standard Model (SM) and beyond. The clean and
readily identifiable signature and large production rate of this process provide an opportunity to precisely
measure the electroweak parameters, constrain the parton distribution functions (PDFs) and also probe
the strong coupling constant $\alpha_s$.  In particular, it is a prominent background in  searches for 
SM processes and physics beyond the SM at the TeV scale~\cite{Blumenschein:2018gtm}.   
Therefore, a precise study of both the inclusive and differential measurements of $Z$ boson plus 
jets production is vital to test the SM and search for new physics (NP). 

Currently, both  the ATLAS and CMS collaborations have reported the measurements of $Z$ boson production associated with zero, one and two jets~\cite{ATLAS:2016bgc,Khachatryan:2016crw,Aaboud:2017hbk,Sirunyan:2018cpw}.  
Although the experimental measurements on $p_T$ and $y$ of jet show very good agreement with theoretical predictions~\cite{Boughezal:2016yfp},
a better theoretical calculation for some other observables (e.g., the total transverse momentum of $Z$ boson and leading jet system) in regions of the phase space dominated by soft/collinear radiation are still needed to reduce the theoretical uncertainties.
Both the fixed-order and resummation techniques could be used to improve the theoretical predictions.  Perturbative QCD corrections to the $Z$ boson plus multijets production at the next-to-leading order (NLO) are widely discussed in literatures~\cite{Arnold:1988dp,Giele:1993dj,Campbell:2002tg,Campbell:2003hd,Berger:2010vm,Ita:2011wn,Campbell:2016tcu,Figueroa:2018chn}. The  NLO effects  from electroweak correction to $Z$ boson plus multijets are also discussed in Refs.~\cite{Kuhn:2005az,Denner:2011vu,Hollik:2015pja,Kallweit:2015dum}.   Beyond the NLO  QCD calculation, the leading threshold logarithms have been  included in Ref.~\cite{Becher:2011fc}. The accuracy to the $Z$ boson plus one jet production has reached to the next-to-next-to-leading order (NNLO)  in QCD interactions
~\cite{Boughezal:2015ded,Ridder:2015dxa,Boughezal:2016isb}.  Recently, the transverse momentum effects from the initial state partons are also discussed in the $Z$ boson plus one jet production~\cite{Deak:2018obv}. 

In this work, we focus on improving the prediction on the kinematical distributions of the  production of the $Z$ boson plus one jet, 
\beq
p+p\to Z(P_Z)+Jet(P_{J})+X \ ,\label{eq1}
\eeq
where $P_Z$ and $P_J$ are the momenta of $Z$ boson and leading jet, respectively. The transverse momentum resummation  ($q_\perp$ resummation) formalism is applied to sum over large logarithm $\ln(Q^2/q_{\perp}^2)$, with $Q\gg q_\perp$, to all orders in the expansion of the strong interaction coupling at the NLO and next-leading logarithm (NLL) accuracy, where $Q$ and $q_\perp$ are the invariant mass and total transverse momentum of $Z$ boson plus leading jet final state system, respectively. The $q_\perp$ resummation technique is based on the transverse momentum dependent (TMD) factorization formalism~\cite{Collins:1981uk,Collins:1981va}, which has been widely discussed in the literature in the color singlet processes, such as Drell-Yan production~\cite{Collins:1984kg}. Extending the $q_\perp$ resummation formalism to processes with more complex color structure have been discussed recently; e.g. heavy quark pair production~\cite{Zhu:2012ts,Li:2013mia,Zhu:2013yxa};  processes involving multijets in the final state~\cite{Sun:2014gfa,Sun:2014lna,Sun:2015doa,Sun:2016kkh,Sun:2016mas,Cao:2018ntd,Sun:2018beb}. Here  we will use the TMD resummation formalism presented in Refs.~\cite{Sun:2014gfa,Sun:2015doa} to discuss the kinematical distributions of $Z$ boson plus one jet. To properly describe the jet in the final state, we should modify $q_\perp$ resummation formalism to include the soft gluon radiation from the final state; see a detailed discussion in Refs.~\cite{Sun:2014gfa,Sun:2014lna,Sun:2015doa,Sun:2016kkh,Sun:2016mas,Cao:2018ntd,Sun:2018beb}. In short, we should resum the large logarithm $\ln(Q^2/q_{\perp}^2)$ when there is  soft gluon radiation outside the observed final-state jet cone.

\noindent{\bf TMD Resummation.}
Our TMD resummation formula can be written as~\cite{Sun:2014lna}:
\begin{align}
&\frac{d^5\sigma}
{dy_Z dy_J d P_{J\perp}^2
d^2\vec{q}_{\perp}}=\sum_{ab}\nn\\
&\left[\int\frac{d^2\vec{b}_\perp}{(2\pi)^2}
e^{-i\vec{q}_\perp\cdot
\vec{b}_\perp}W_{ab\to ZJ}(x_1,x_2,b_\perp)+Y_{ab\to ZJ}\right] \ ,\label{resumy}
\end{align}
where $y_Z$ and $y_J$ denote the rapidity of the Z boson and the leading jet;
$P_{J\perp} (P_{Z\perp})$  and $\vec{q}_\perp=\vec{P}_{Z\perp}+\vec{P}_{J\perp}$ are the leading jet  ($Z$ boson) transverse momentum
 and  the imbalance transverse momentum of the Z boson and the jet system.
The first term ($W_{ab\to ZJ}$) contains all order resummation effect
and the second term ($Y_{ab\to ZJ}$) accounts for the difference between
the fixed order result and the so-called asymptotic
result which is given by expanding the resummation result
to the same order in $\alpha_s$ as the fixed order term. $x_1$ and $x_2$ are the momentum fractions of
the incoming hadrons carried by the two incoming partons, 
\begin{eqnarray}
x_{1,2}=\frac{\sqrt{m_Z^2+P^2_{Z\perp}}e^{\pm y_Z}+\sqrt{P^2_{J\perp}}e^{\pm y_J}}{\sqrt{S}} \ ,
\end{eqnarray}
where $m_Z$ and $S$ are the $Z$ boson mass and squared collider energy, respectively.
The all order resummation result  $W_{ab\to ZJ}$ can be further written as,
\begin{align}
&W_{ab\to ZJ}\left(x_1,x_2,b\right)\nn\\
&=x_1f_a(x_1,\mu_F=b_0/b_\perp) x_2f_{b}(x_2,\mu_F=b_0/b_\perp)\nn\\
&\times {H}_{ab\to ZJ} (s, \mu_{\rm res},\mu_R)
 e^{-S_{\rm Sud}(s,\mu_{\rm res},b_\perp)}e^{-\mathcal{F}_{NP}},  \label{resum}
\end{align}
where $s=x_1x_2S$, $b_0=2e^{-\gamma_E}$ with $\gamma_E$ being the Euler constant, $\mu_{\rm res}$ is the
resummation scale to apply the TMD factorization in the resummation calculation. $\mu_{\rm res}$ is
also the scale to define the TMDs in the Collins 2011 scheme~\cite{Collins:2011zzd}. $\mu_R$ is the renormalization scale.
$f_{a,b}(x,\mu_F)$ are the PDFs for the incoming
partons $a$ and $b$,  $\mu_F$ is factorization scale of the PDFs and $b_\perp=b/\sqrt{1+b^2/b_{\rm max}^2}$
with $b_{\rm max}=1.5~{\rm GeV}^{-1}$, which is introduced to factor out the non-perturbative contribution $e^{\mathcal{F}_{NP}}$, arising from the large $b$ region (with $b\gg b_\perp$)~\cite{Landry:1999an,Landry:2002ix,Sun:2012vc,Su:2014wpa},
\begin{equation}
\mathcal{F}_{NP}(Q^2,\textbf{b})=g_1b^2+g_2\ln\dfrac{Q}{Q_0}\ln\dfrac{b}{b_\perp},
\end{equation}
where $g_1=0.21$, $g_2=0.84$ and $Q_0^2=2.4~{\rm GeV}^2$~\cite{Su:2014wpa}.

The Sudakov form factor can be expressed as,
\begin{eqnarray}
S_{\rm Sud}=\int^{\mu_{\rm res}^2}_{b_0^2/b_\perp^2}\frac{d\mu^2}{\mu^2}
\left[\ln\left(\frac{s}{\mu^2}\right)A+B_1+B_2+D\ln\frac{1}{R^2}\right]\ ,\nn\\ 
\label{su}
\end{eqnarray}
where $R$ denotes the jet cone size of the final state jet.
The coefficients $A$, $B_{1,2}$ and $D$ can be expanded
perturbatively in $\alpha_s$, which is  $g_s^2/(4\pi)$.
\beq
A/B_{1,2}/D=\sum_{n=1}^\infty \left(\dfrac{\alpha_s}{\pi}\right)^nA^{(n)}/B_{1,2}^{(n)}/D^{(n)}.
\eeq
For $q\bar{q}\to Zg$ channel, 
we have 
\begin{align}
A^{(1)}&=C_F, & A^{(2)} &=\dfrac{1}{2}C_F K, & B_1^{(1)}&=-\frac{3}{2}C_F,\nn\\
B_2^{(1)}&=0, & D^{(1)}&=\frac{1}{2}C_A.\nn\\
\end{align}
For $gq\to Zq$ channel, we have
\begin{align}
A^{(1)}&=\frac{1}{2}(C_F+C_A) , &, A^{(2)}&=\dfrac{1}{2}\dfrac{C_F+C_A}{2}K,\nn\\
 B_1^{(1)}&=(-C_A\beta_0-\dfrac{3}{4}C_F),&
B_2^{(1)}&=\frac{1}{2}(C_F-C_A)\ln\left(\dfrac{u}{t}\right), \nn\\
D^{(1)}&=\frac{1}{2}C_F,
\end{align}
where $C_F=\dfrac{4}{3}$,  $C_A=3$ and $K=\dfrac{67}{18}-\dfrac{\pi^2}{6}C_A-\dfrac{5}{9}N_f$;  $\beta_0=(11-2/3N_f)/12$ with $N_f=5$ being the number of effective light quarks.
Here $t=(P_a-P_Z)^2$ and $u=(P_a-P_J)^2$ with the incoming parton momentum $P_a$. They are the usual Mandelstam variables for the partonic $2\to 2$ process.
The coefficients $A$ and $B_1$ come from the energy evolution effect in the TMD PDFs ~\cite{Ji:2004wu}, so that they only depend on the flavor of the incoming partons and are independent of the scattering processes. The  coefficient $B_2$ describes the soft gluon interaction between initial and final states.  The factor $D$ quantifies the effect of soft gluon radiation which goes outside the jet cone, hence it depends on the jet cone size $R$.  Furthermore, the narrow jet approximation~\cite{Jager:2004jh,Mukherjee:2012uz} is applied to simplify the calculation, and we only keep the term proportional to $\ln(1/R^2)$. In our numerical calculation, the  $A^{(2)}$ terms will also be included in our analysis since it is associated with the incoming parton distribution and universal for all processes~\cite{Catani:2000vq}.

By applying the TMD factorization with Collins 2011 scheme, we obtain the hard
factor $H_{q\bar{q}\to Zg}$ in Eq.~(\ref{resum}), at the one-loop order, as
\begin{widetext}
\begin{eqnarray}
H^{(1)}_{q\bar{q}\to Zg} &=& H_{q\bar{q}\to Zg}^{(0)}\dfrac{\alpha_s}{2\pi}\left\lbrace\left[-2 \beta _0 \ln \left(\frac{R^2 P_J^2}{\mu _{\text{res}}^2}\right)+\frac{1}{2} \ln ^2\left(\frac{R^2
   P_J^2}{\mu
   _{\text{res}}^2}\right)+\text{Li}_2\left(\frac{m_Z^2}{m_Z^2-t}\right)+\text{Li}_2\left(\frac{m_Z^2}{m_Z^2-u}\right)\nonumber
\right.\right.\\
&-&\ln
   \left(\frac{\mu _{\text{res}}^2}{m_Z^2}\right) \ln \left(\frac{s m_Z^2}{t u}\right)-\frac{1}{2} \ln ^2\left(\frac{\mu
   _{\text{res}}^2}{m_Z^2}\right)+\frac{1}{2} \ln ^2\left(\frac{s}{m_Z^2}\right)-\frac{1}{2} \ln ^2\left(\frac{t
   u}{m_Z^4}\right)+\ln \left(\frac{t}{m_Z^2}\right) \ln \left(\frac{u}{m_Z^2}\right)\nonumber\\
&+&\left.\frac{1}{2} \ln
   ^2\left(\frac{m_Z^2-t}{m_Z^2}\right)+\frac{1}{2} \ln^2 \left(\frac{m_Z^2-u}{m_Z^2}\right)-\frac{1}{2} \ln
   ^2\left(\frac{1}{R^2}\right)-\frac{2 \pi ^2}{3}+\frac{67}{9} -\frac{23 N_f}{54} \right]C_A+6\beta_0\ln\dfrac{\mu_R^2}{\mu_{\rm res}^2}\nonumber\\
&+&\left.\left[ 2 \ln \left(\frac{s}{m_Z^2}\right) \ln \left(\frac{\mu _{\text{res}}^2}{m_Z^2}\right)-\ln ^2\left(\frac{\mu
   _{\text{res}}^2}{m_Z^2}\right)-3 \ln \left(\frac{\mu _{\text{res}}^2}{m_Z^2}\right)-\ln ^2\left(\frac{s}{m_Z^2}\right)+\pi
   ^2-8\right]C_F\right\rbrace+\delta H^{(1)}_{q\bar{q}\to Zg},\nonumber\\
\end{eqnarray}
\end{widetext}
The leading order matrix element for $q\bar{q}\to Zg$ is,
\begin{align}
H_{q\bar{q}\to Zg}^{(0)}&=\dfrac{8\pi}{3}\alpha_sC_F\left(g_V^2+g_A^2\right)\left[\dfrac{u}{t}+\dfrac{t}{u}+\dfrac{2m_Z^2(m_Z^2-u-t)}{tu}\right].
\end{align}
The vector and axial-vector gauge couplings between $Z$ boson and quarks are,
\begin{equation}
g_V=\dfrac{g_W}{2\cos\theta_W}(\tau_3^q-2Q_q \sin\theta_W^2),\quad g_A=\dfrac{g_W}{2\cos\theta_W}\tau_3^q,
\end{equation}
where $g_W$ and $\theta_W$ are the weak gauge coupling and weak mixing angle, respectively. $\tau^3_q$ is the third component of the quark weak isospin and $Q_q$ is the electric charge of quark. $\delta H^{(1)}$ represents terms  which are not proportional to $H^{(0)}$ and can be found in Ref.~\cite{Arnold:1988dp}.
Similarly, for the subprocess $g+q\rightarrow Z+q$, we have
\begin{widetext}
\begin{eqnarray}
H^{(1)}_{qg\to Zq} &=& H_{qg\to Zq}^{(0)}\dfrac{\alpha_s}{2\pi}\left\lbrace\left[-\dfrac{3}{2} \ln \left(\frac{R^2 P_J^2}{\mu _{\text{res}}^2}\right)+\frac{1}{2} \ln ^2\left(\frac{R^2
   P_J^2}{\mu _{\text{res}}^2}\right)+2\ln\left(\frac{u}{m_Z^2}\right) \ln \left(\frac{\mu_{\rm res}^2}{m_Z^2}\right)\nonumber
\right.\right.\nonumber\\
&-&\left. \ln ^2\left(\frac{\mu _{\text{res}}^2}{m_Z^2}\right)-3 \ln \left(\frac{\mu _{\text{res}}^2}{m_Z^2}\right)- \ln ^2\left(\frac{u}{m_Z^2}\right)-\frac{1}{2} \ln ^2\left(\frac{1}{R^2}\right)-\dfrac{2\pi^2}{3}-\dfrac{3}{2}\nonumber \right]C_F+6\beta_0\ln\dfrac{\mu_R^2}{\mu_{\rm res}^2}\nonumber\\
&+&\left[
-\text{Li}_2\left(\frac{m_Z^2}{s}\right)+\text{Li}_2\left(\frac{m_Z^2}{m_Z^2-t}\right)-\ln \left(\frac{\mu
   _{\text{res}}^2}{m_Z^2}\right) \ln \left(\frac{u m_Z^2}{s t}\right)-\frac{1}{2} \ln ^2\left(\frac{\mu
   _{\text{res}}^2}{m_Z^2}\right)
   \right.\nonumber\\
&-&\frac{1}{2} \ln ^2\left(\frac{s t}{m_Z^4}\right)+\ln \left(\frac{s}{m_Z^2}\right) \ln
   \left(\frac{t}{m_Z^2}\right)-\ln \left(\frac{s}{m_Z^2}\right) \ln \left(\frac{t}{m_Z^2-s}\right)-\frac{1}{2} \ln
   ^2\left(\frac{s}{m_Z^2}\right)\nonumber\\
&+& \left.\left.\frac{1}{2} \ln ^2\left(\frac{m_Z^2-t}{m_Z^2}\right)+\frac{1}{2} \ln
   ^2\left(\frac{u}{m_Z^2}\right)+\frac{\pi ^2}{2}
\right]C_A\right\rbrace+\delta H^{(1)}_{qg\to Zq},\nonumber\\
\end{eqnarray}
\end{widetext}
where the leading order matrix element is,
\beq
H_{qg\to Zq}^{(0)}=-\pi\alpha_sC_F\left(g_V^2+g_A^2\right)\left[\dfrac{s}{t}+\dfrac{t}{s}+\dfrac{2m_Z^2(m_Z^2-s-t)}{ts}\right].
\eeq

We should note that the non-global logarithms (NGLs) could also contribute to this process.
The NGLs arise from some special kinematics of two soft gluon radiations, in which the first one is radiated outside of the jet which subsequently radiates a second gluon into the jet~\cite{Dasgupta:2001sh,Dasgupta:2002bw,Banfi:2003jj,Forshaw:2006fk}.
Recently, the NGLs effects were studied in Ref.~\cite{Chien:2019gyf} in the framework of  soft-collinear effective theory and it shows that their contributions are negligible  when $P_{J\perp}>30~{\rm GeV}$.
Therefore  we will not consider the NGLs in the following numerical calculations. The additional resummation effect of $\ln R$ is beyond the scope of this paper and has also been discussed  in Ref.~\cite{Chien:2019gyf}. 

\noindent{\bf $Z$ Boson Plus Jet Production at the LHC.}
We apply the resummation formula of Eq.~(\ref{resumy}) to calculate
the differential and total cross sections of the Z boson
production associated with a high energy jet.  The anti-$k_t$ jet  algorithm  with jet cone size $R=0.4$  will be used to define the observed jet as discussed in Refs.~\cite{Mukherjee:2012uz,Sun:2015doa}.

Before we present our numeric results, we would like to comment on the
cross-check of our resummation method.  We perform the fixed order expansion of
the integral of Eq.~(\ref{resumy}) to obtain the total cross section, and  compare it  with the
fixed order prediction. The $Y$-term is vanishing when $q_\perp$ goes to zero in the 
resummation framework, thus the cross section in the small $q_\perp$ region (from $q_\perp=0$ to a small value $q_{\perp,0}$, about 1 GeV)  can be obtained by integrating the distribution  of the asymptotic part and the one-loop virtual diagram contribution.  The cross section in the large $q_\perp$
region  ($q_\perp>q_{\perp,0}$) is infrared safe and can be numerically calculated directly. Thus, the total cross section can be written as~\cite{Balazs:1997xd},
\beq
\sigma_{NLO}=\int_0^{q_{\perp,0}^2}dq_{\perp}^2\dfrac{d\sigma_{NLO}^{virtual+real}}{dq_{\perp}^2}+\int_{q_{\perp,0}^2}^{\infty}dq_{\perp}^2\dfrac{d\sigma_{NLO}^{real}}{dq_{\perp}^2}.\nn\\
\eeq
Numerically, we find that the above procedure reproduces the NLO cross sections from MCFM~\cite{Campbell:2015qma} with slight difference, ranging from 2\% for $R=0.4$ to $0.2\%$ for $R=0.2$. Clearly, this discrepancy arises from the narrow jet approximation made in our derivations. Following the procedure of Ref.~\cite{Sun:2016kkh}, we parameterize this difference as function of $R$: $H^{(0)}\dfrac{\alpha_s}{2\pi}(0.74R-6.44R^2)$ for the range of $0.2<R<0.6$, which will be considered as part of our NLO contribution $H^{(1)}$.  

\begin{figure*}
\includegraphics[width=0.23\textwidth]{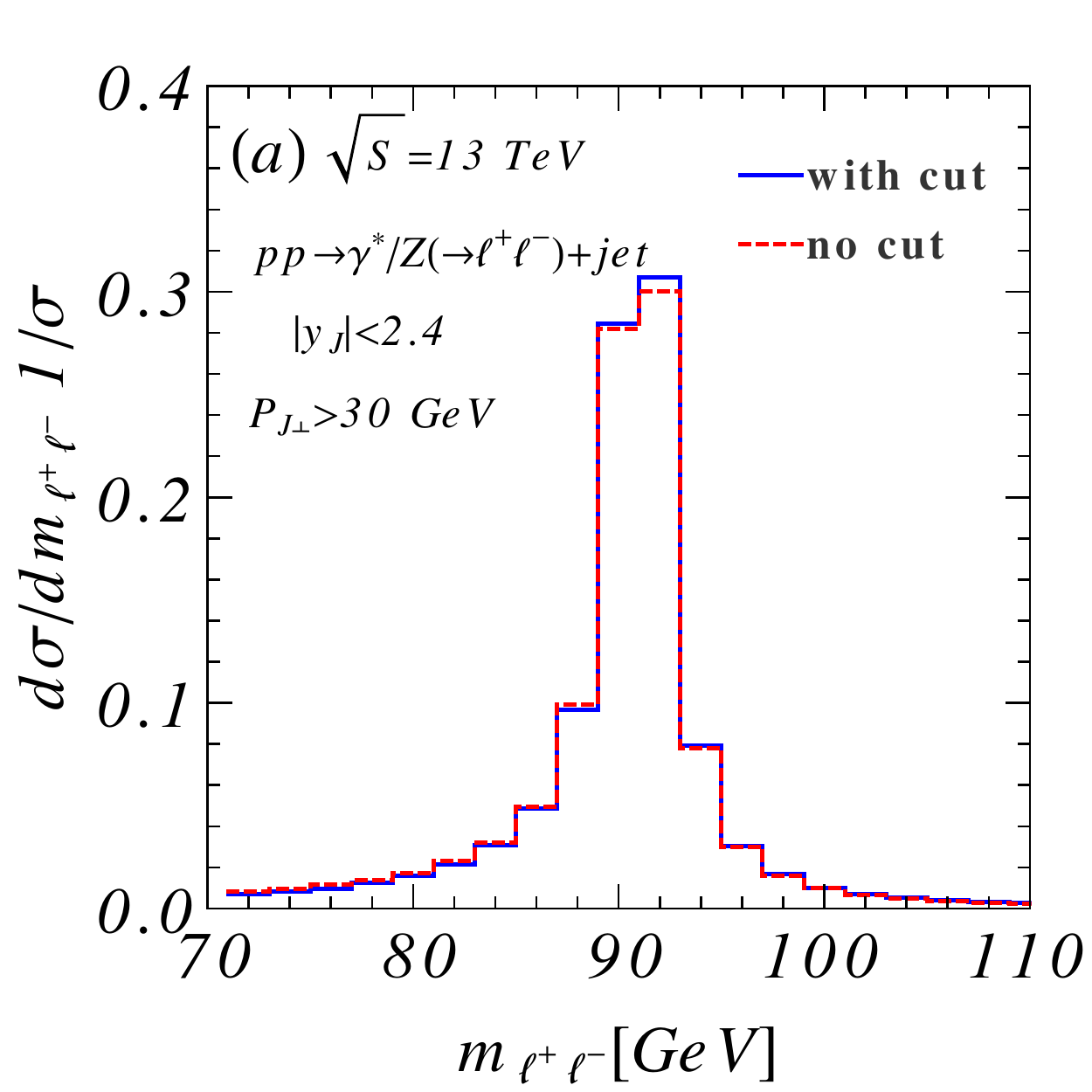}
\includegraphics[width=0.23\textwidth]{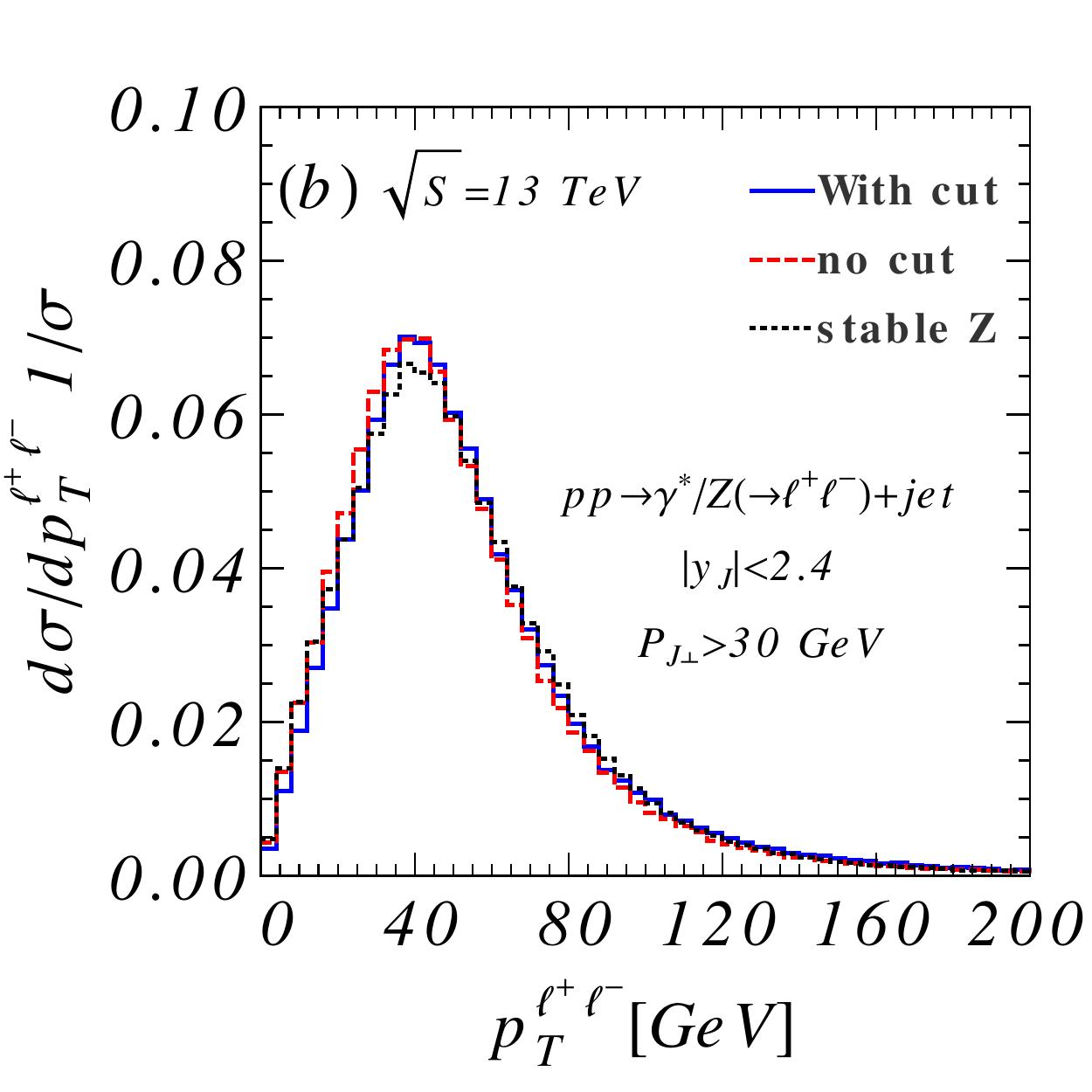}
\includegraphics[width=0.23\textwidth]{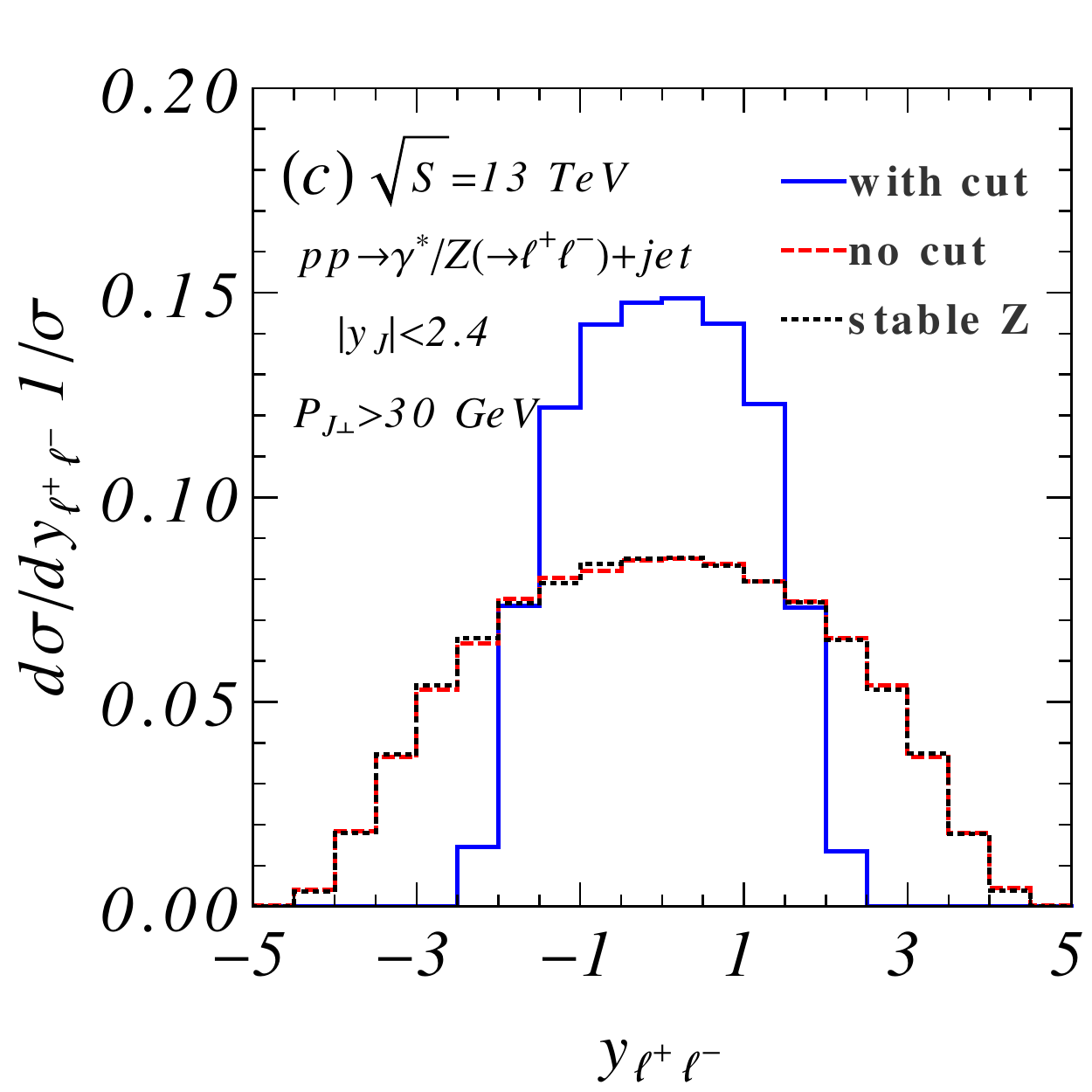}
\includegraphics[width=0.23\textwidth]{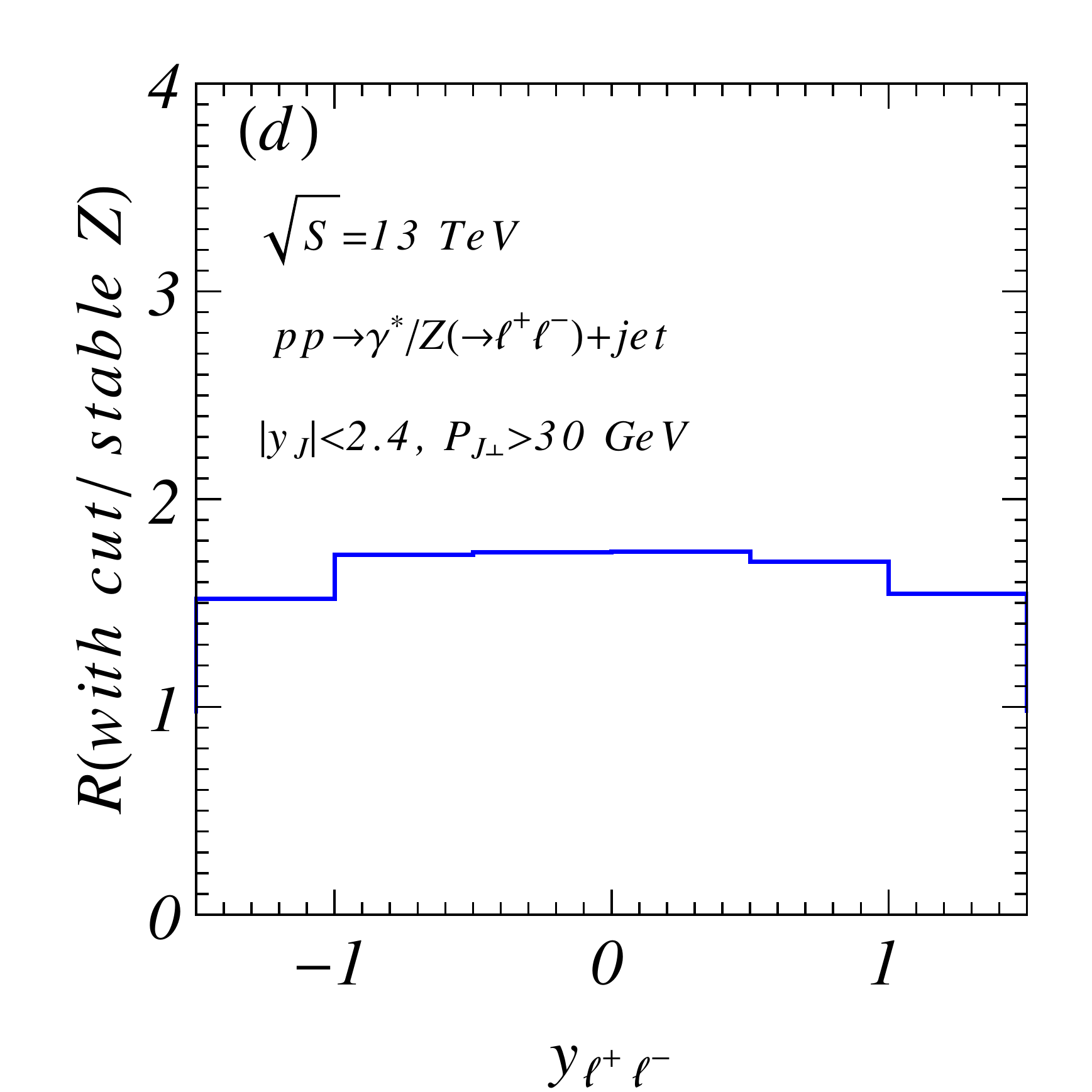}
\caption{Comparison of normalized differential distributions of the (a) invariant mass; (b) transverse momentum; and (c) rapidity; of the lepton pairs predicted by PYTHIA8  	
for the $Z$-boson plus jet production at the $\sqrt{S}=13~{\rm TeV}$ LHC with  $|y_J|<2.4$ and $P_{J\perp}>30~{\rm GeV}$. 
The blue sold lines show the distributions with the kinematic cuts imposed on the leptons, as done in the CMS measurement~\cite{Sirunyan:2018cpw}, which are 
$71~{\rm GeV} < m_{\ell^+\ell^-}< 111~{\rm GeV}, \quad p^{\ell^{\pm}}_T>20~{\rm GeV}, {\rm and} \quad |\eta_{\ell^{\pm}}|<2.4$.
The red dashed lines show the predictions without imposing the kinematic cuts on the decay leptons. The black dotted lines show the predictions for a stable $Z$ boson production,
hence,  $p_T^{\ell^+\ell^-}\equiv p_T^Z$ and $y_{\ell^+\ell^-}\equiv y_Z$. 
(d) the ratio of normalized rapidity distribution between with cut (blue  solid line in (c)) and stable $Z$ boson (black dashed line in (c)).
\label{fig:lldis} }
\end{figure*}%

\begin{figure}
\centering
\includegraphics[width=0.23\textwidth]{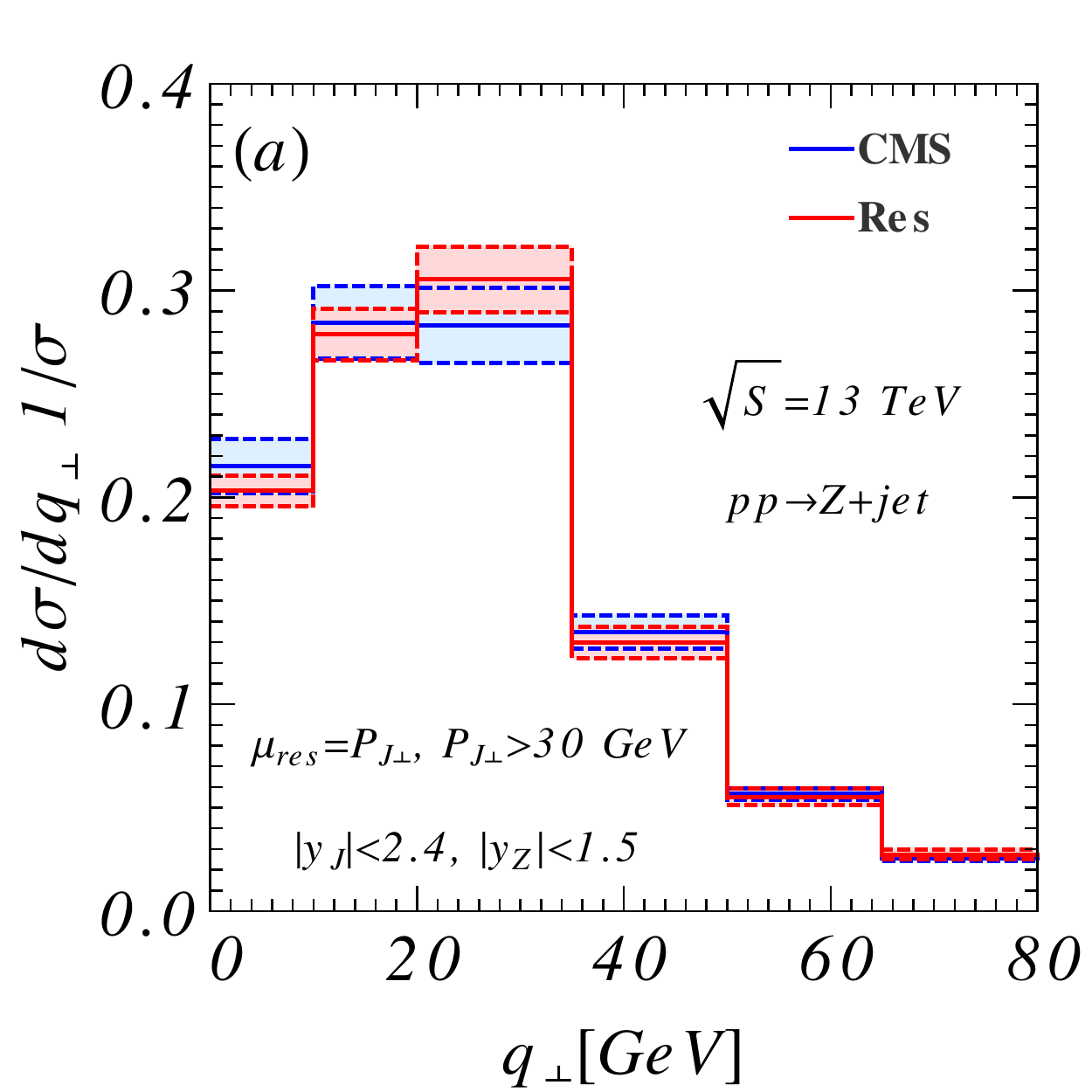}
\includegraphics[width=0.23\textwidth]{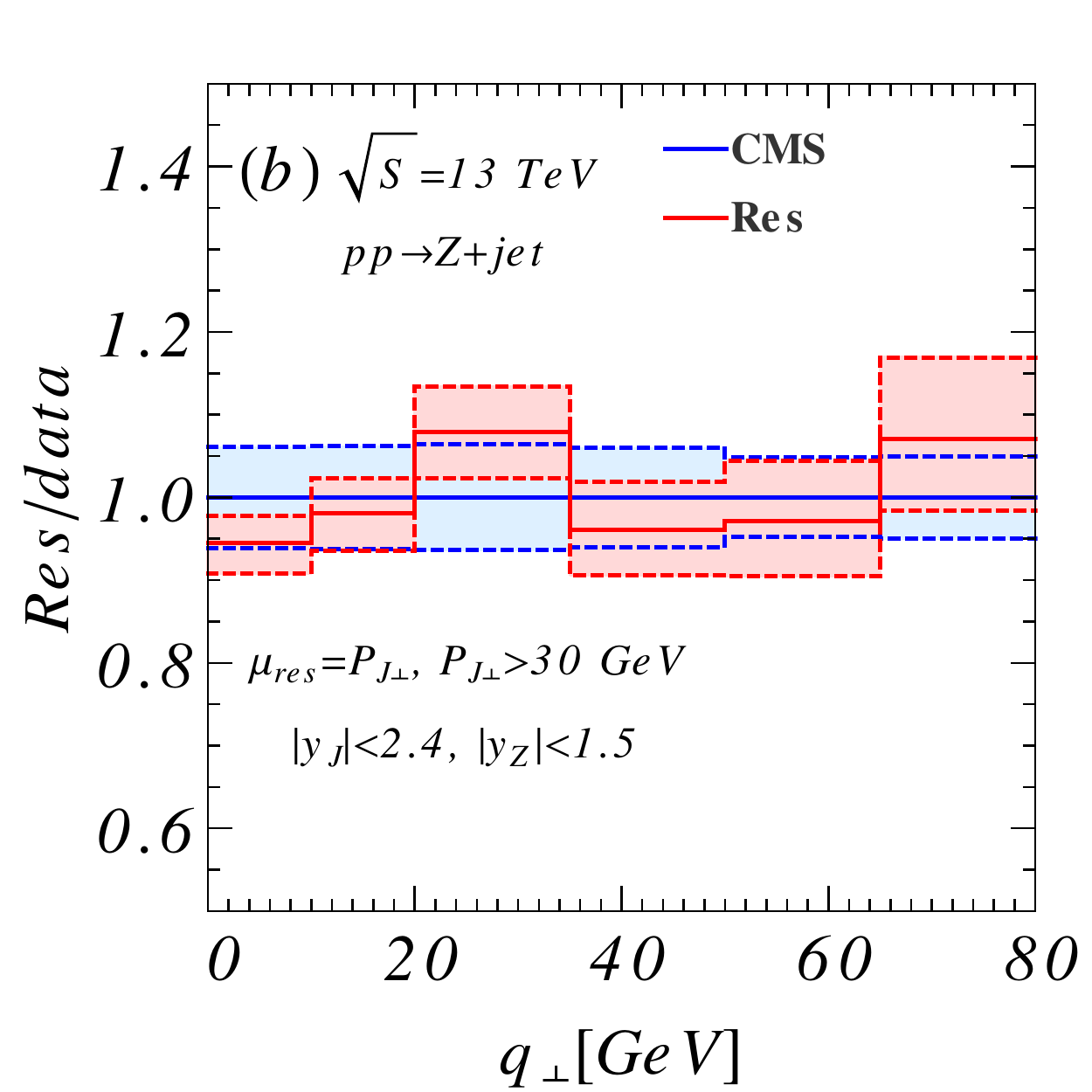}
\caption{(a) The normalized $q_\perp$ distribution of the $Z$ boson plus one jet system, produced at the $\sqrt{S}=13~{\rm TeV}$ LHC with $|y_J|<2.4$ and $P_{J\perp}>30~{\rm GeV}$. The blue and red bands represent the CMS experimental uncertainty~\cite{Sirunyan:2018cpw} and the resummation calculation (Res) scale uncertainty, respectively. (b) The ratio of resummation prediction to CMS data as a function of $q_\perp$.  }
\label{fig:qt}
\end{figure}

Recently, CMS collaboration has reported the measurement of the $q_{\perp}$ spectrum of $Z$ boson plus  jet  production at the 13 TeV LHC~\cite{Sirunyan:2018cpw}. 
Since the experimental measurement was done with certain kinematic cuts imposed on the final state leptons, its result cannot be directly compared to the current theory prediction which is for an on-shell $Z$ boson production associated with one or more high-$P_{J\perp}$ jets.
In order to compare to this data, we need to estimate the effect of those kinematic cuts to our theory prediction. 
This estimation can be done by employing a reweighting procedure based on the PYTHIA8 simulations. 
For example, the differential cross section of the imbalance transverse momentum of the $Z$ boson and  leading jet system ($q_\perp$), after imposing the kinematic cuts on the decay leptons of the $Z$ boson, can be written as
\beq
\dfrac{d\sigma}{dq_\perp}\bigg|_{\rm decay}=\dfrac{d\sigma}{dq_\perp}\bigg|_{\rm {stable}, Z}\times \kappa(m_{\ell^+\ell^-},y_{\ell^+\ell^-},p^{\ell^+\ell^-}_T),
\eeq
where $\kappa(m_{\ell^+\ell^-},p^{\ell^+\ell^-}_T,y_{\ell^+\ell^-})$ is the reweighting factor which depends  on lepton pair invariant mass ($m_{\ell^+\ell^-}$), transverse momentum  ($p^{\ell^+\ell^-}_T$) and rapidity ($y_{\ell^+\ell^-}$).  $d\sigma/dq_\perp |_{\rm {stable}, Z}$ is the differential cross section with stable $Z$ boson production.  Figure~\ref{fig:lldis} shows the normalized $m_{\ell^+\ell^-}$,  $p^{\ell^+\ell^-}_T$ and $y_{\ell^+\ell^-}$ distributions at the $\sqrt{S}=13~{\rm TeV}$, with  $|y_J|<2.4$ and $P_{J\perp}>30~{\rm GeV}$, as predicted by the Monte Carlo event generator PYTHIA8~\cite{Sjostrand:2007gs}. The blue solid lines show the distributions after we impose the following kinematic cuts on the leptons (labelled as `with cut')~\cite{Sirunyan:2018cpw},
\begin{eqnarray}
& 71~{\rm GeV} < m_{\ell^+\ell^-}< 111~{\rm GeV}, \nonumber\\
&\quad p^{\ell^{\pm}}_T>20~{\rm GeV}, {\rm and} \quad |\eta_{\ell^{\pm}}|<2.4.
\label{eq:cuts}
\end{eqnarray}
The red dashed lines show the prediction of $pp\to \gamma^*/Z(\to \ell^+\ell^-)+jet$ without the above kinematic cuts imposed on the $Z$-decay leptons (labelled as `no cut'), while the black dotted lines show the prediction with stable $Z$ boson production (labelled as `stable $Z$'). 
It is clear that the normalized distributions of 
$m_{\ell^+\ell^-}$ and $p_T^{\ell^+\ell^-}$
are not sensitive to the imposed lepton kinematic cuts. 
On the contrary, the kinematic cuts on the decay leptons significantly modified the shape of the rapidity distribution of the lepton pairs, cf. Fig.~\ref{fig:lldis}(c).  Therefore, to a very good approximation, we can assume that the reweighting factor $\kappa$ only depends on the value of $y_{\ell^+\ell^-}$, i.e.
\beq
\kappa(m_{\ell^+\ell^-},p^{\ell^+\ell^-}_T,y_{\ell^+\ell^-})\simeq \kappa(y_{\ell^+\ell^-}).
\eeq
The kinematic cuts imposed on the leptons, as in Eq.~(\ref{eq:cuts}), will constrain the allowed rapidity range of the lepton pair, and approximately  $|y_{\ell^+\ell^-}|<1.5$. 
 Figure ~\ref{fig:lldis}(d) shows the ratio of normalized rapidity distribution between with cut and stable $Z$ boson prediction in Fig. ~\ref{fig:lldis}(c). It is clear that $\kappa(y_{\ell^+\ell^-})$ does not strongly depend on $y_{\ell^+\ell^-}$ for $|y_Z|<1.5$.
we could approximate a constant reweighting factor to describe the effect of the kinematic cuts on the $Z$-decay leptons; i.e. 
\beq
\kappa(y_{\ell^+\ell^-})\simeq \kappa.
\eeq
Although $\kappa$ is estimated based on LO prediction given by the PYTHIA8 event generaotr, the theoretical uncertainties from higher order corrections are not significant~\cite{Boughezal:2016yfp}.
Therefore, under this approximation, we have 
\beq
\dfrac{d\sigma}{ dq_\perp}\bigg|_{\rm decay}
\simeq  \kappa \times 
\bigg(\dfrac{d\sigma}{ dq_\perp}\bigg|_{\rm {stable}, Z}\bigg)  , ~ \rm{for}~ |y_{\ell^+\ell^-}|<1.5,
\eeq
with the kinematic cuts imposed in the  CMS measurement~\cite{Sirunyan:2018cpw}. 
For the normalized distribution, the $\kappa$ dependence would 
be cancel out and yield  the following relations:
\beq
\dfrac{d\sigma}{ \sigma dq_\perp}\bigg|_{\rm decay}
\simeq 
\bigg(\dfrac{d\sigma}{ \sigma dq_\perp}\bigg|_{\rm {stable}, Z}\bigg)  , ~ \rm{for}~ |y_{\ell^+\ell^-}|<1.5.
\eeq
This approximation is expected to hold well better than the theoretical uncertainty of the normalized $q_\perp$ differential cross section which is at the order of 10\%, cf. Fig.~\ref{fig:qt}. Hence, the small correction arising from taking into account the full rapidity dependence of the re-weighting factor can be ignored in this study.

\begin{figure}
	\centering
	\includegraphics[width=0.23\textwidth]{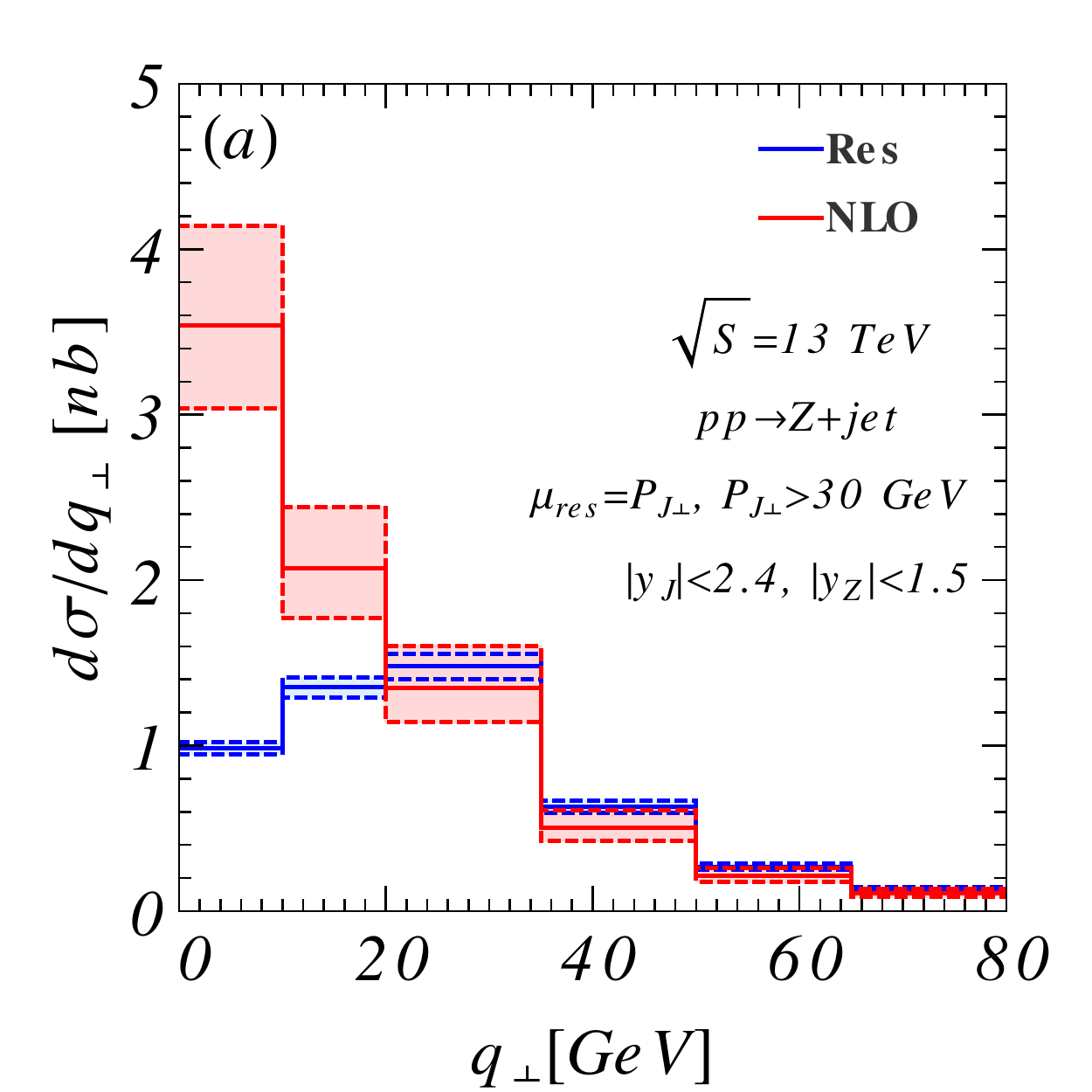}
	\includegraphics[width=0.23\textwidth]{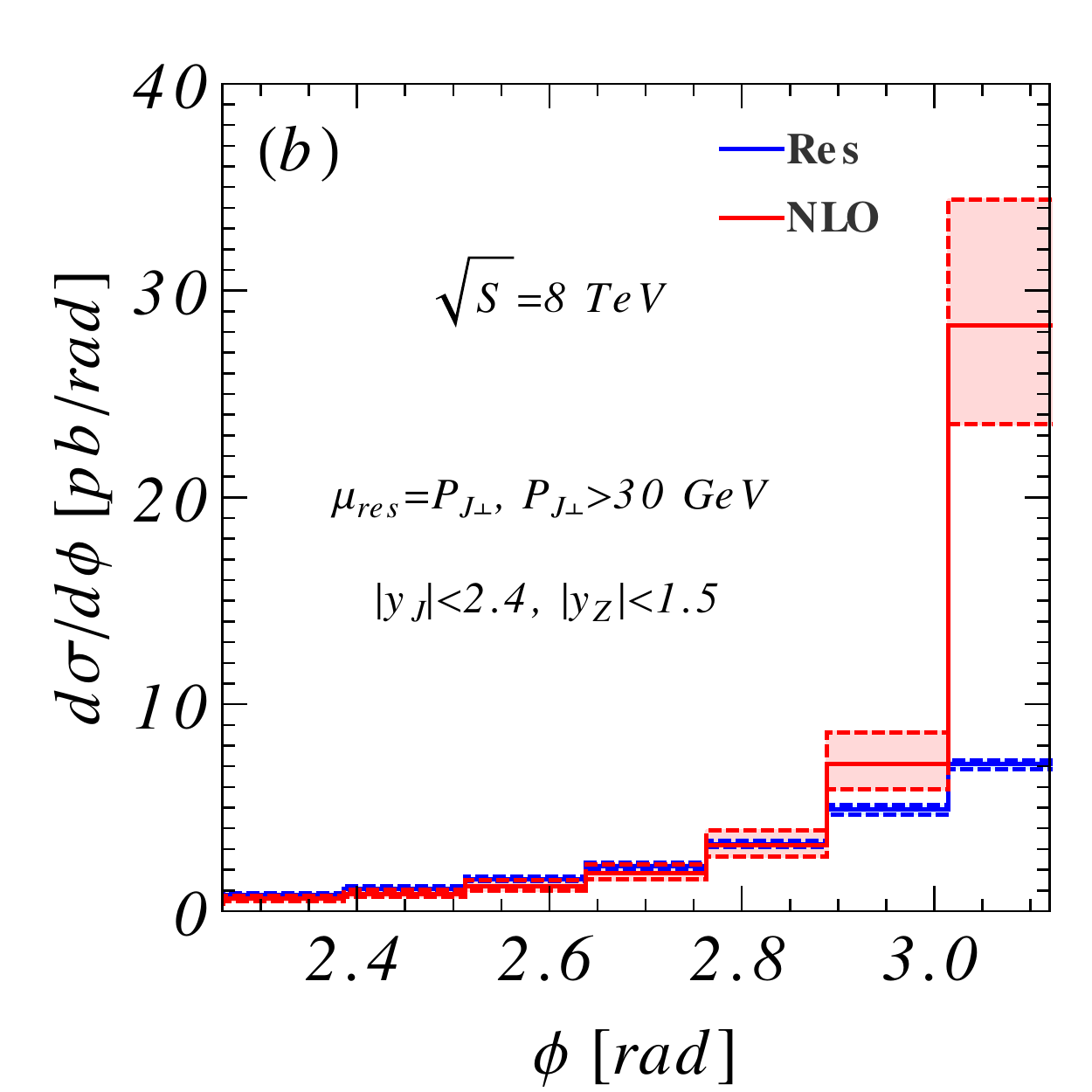}
	\caption{The $q_\perp$ (a) and $\phi$ (b) distributions from resummation calculation (blue band) and NLO prediction (red band) at the $\sqrt{S}=13~{\rm TeV}$ and $\sqrt{S}=8~{\rm TeV}$ LHC, with  $|y_J|<2.4$ and $P_{J\perp}>30~{\rm GeV}$, respectively.}
	\label{fig:qtF}
\end{figure}

\begin{figure}
\centering
\includegraphics[width=0.23\textwidth]{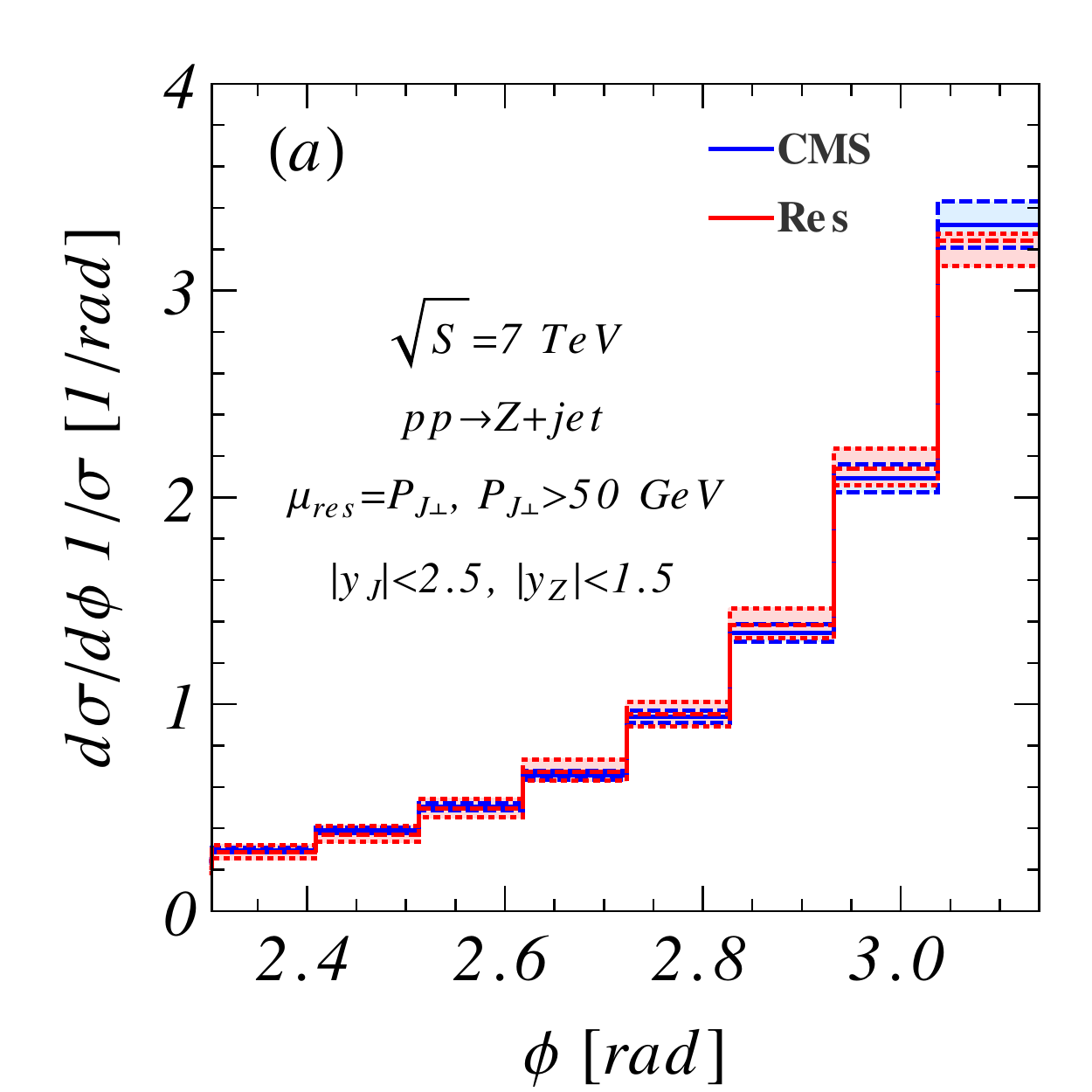}
\includegraphics[width=0.23\textwidth]{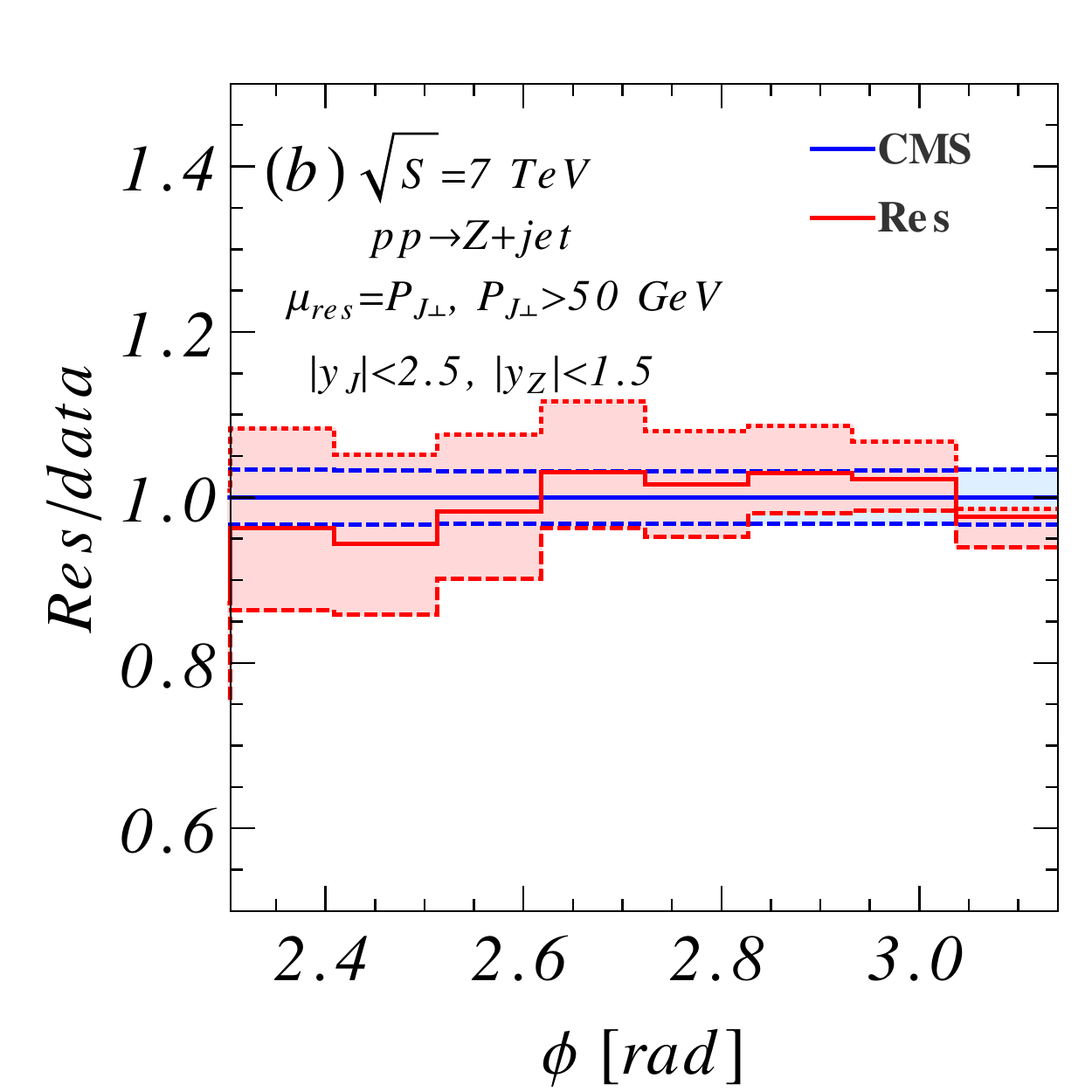}
\caption{The normalized distribution of $\phi$, the azimuthal angle between the final state jet and $Z$ boson measured in the laboratory frame, for $pp\to Z+jet$ production at the $\sqrt{S}=7~{\rm TeV}$ LHC with $|y_J|<2.5$, $|y_Z|<1.5$ and $P_{J\perp}>50~{\rm GeV}$. The blue and red bands represent the CMS experimental uncertainty~\cite{Chatrchyan:2013tna} and the resummation calculation (Res) scale uncertainty, respectively.}
\label{fig:phi7}
\end{figure}

\begin{figure}
\centering
\includegraphics[width=0.23\textwidth]{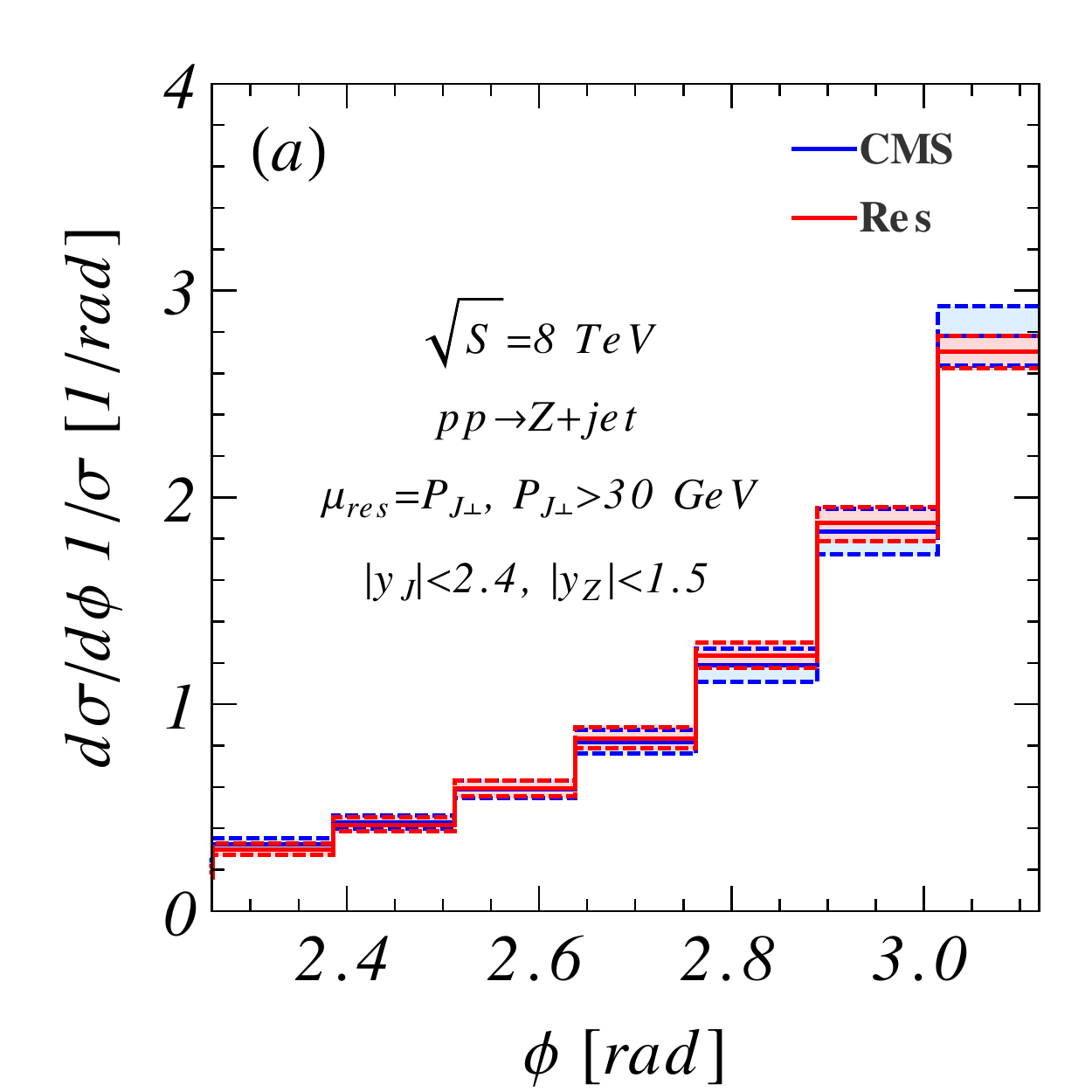}
\includegraphics[width=0.23\textwidth]{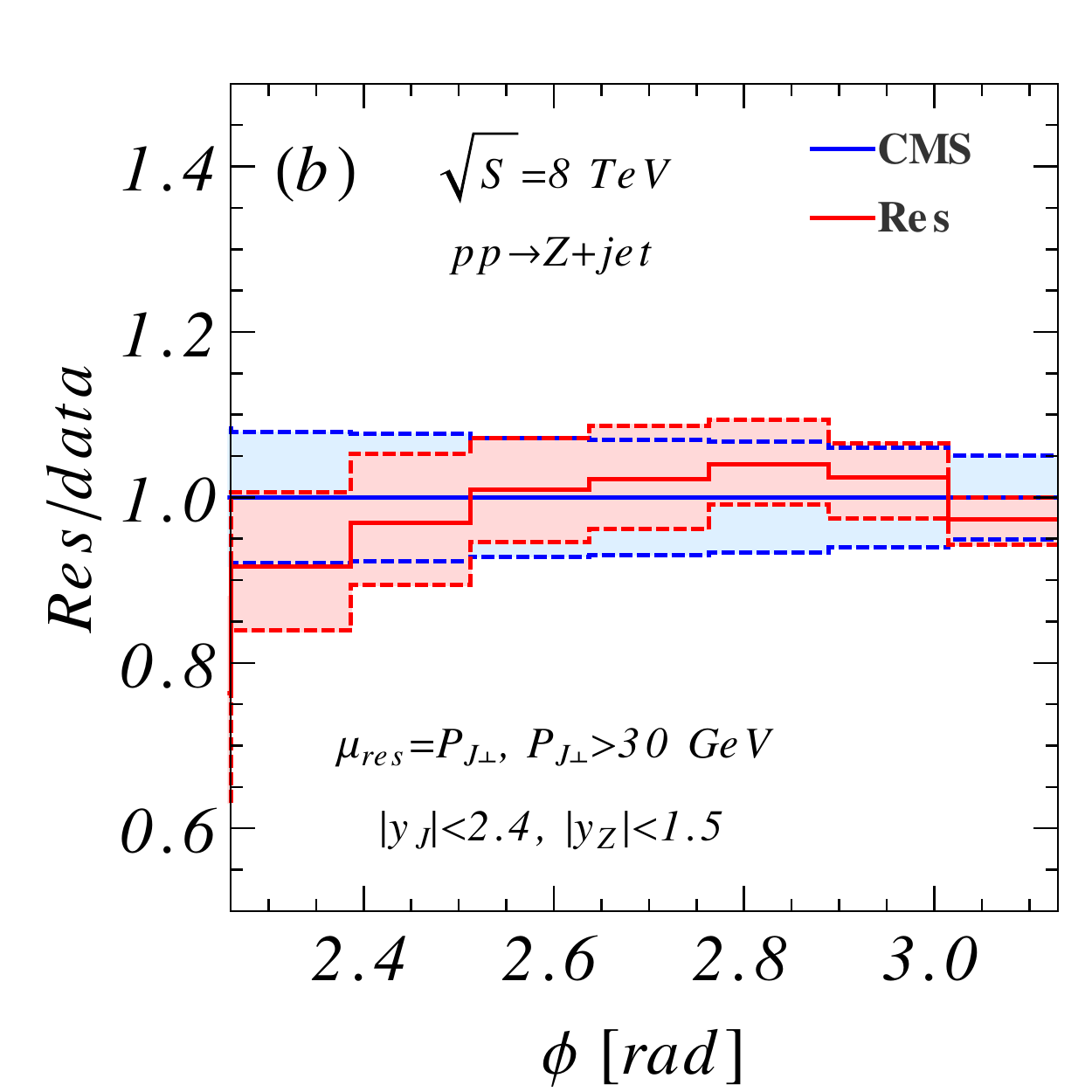}
\caption{Same as Fig.~\ref{fig:phi7}, but for the 8 TeV LHC~\cite{Khachatryan:2016crw}.}
\label{fig:phi8}
\end{figure}

We calculate the normalized $q_\perp$ distribution of $Z$ boson plus one jet production at the $\sqrt{S}=13~{\rm TeV}$ LHC with CT14NNLO PDF~\cite{Dulat:2015mca}, after imposing the kinematic cuts with $|y_J|<2.4$ and $P_{J\perp}>30~{\rm GeV}$, and the results are shown in Fig.~\ref{fig:qt}(a). We fix the resummation scale $\mu_{\rm res}=P_{J\perp}$, while the renormalization scale $\mu_{\rm R}$ is taken to be $H_T=\sqrt{m_Z^2+P_{J\perp}^2}$.  The factorization scale $\mu_F$ of the $Y$-term is also taken to be $H_T$. 
We estimate the scale uncertainties in our calculation by simultaneously varying the scales  $\mu_{\rm R}$  and $\mu_F$ by a factor of two around the central value $H_T$ with a correlated  way. The blue and red bands represent the experimental uncertainty and scale uncertainty, respectively.  In Fig.~\ref{fig:qt}(b), we compare the predictions from our resummation calculation to the CMS data by taking the ratio of their $q_\perp$ differential distributions. It is clear that our resummation calculation agree well with the experimental data.
We also show the comparision between resummation calculation and NLO prediction in Fig.~\ref{fig:qtF}(a). It is clear that there is a large deviation between NLO and resummation calculation in the  small $q_\perp$ region.

The azimuthal angle  ($\phi$) between the final state jet and $Z$ boson measured in the laboratory frame is related to the $q_\perp$ distribution, and is thus  sensitive to the soft gluon radiation. The advantage of studying the $\phi$ distribution is that it only depends on the moving directions of the final state jet and $Z$ boson.  This observable was measured by the CMS Collaboration at the 7 and 8 TeV LHC~\cite{Chatrchyan:2013tna,Khachatryan:2016crw}.
In Figs.~\ref{fig:phi7} and~\ref{fig:phi8}, we compare the normalized $\phi$ angle distribution at the 7 TeV and 8 TeV LHC, respectively. 
Similar to the $q_\perp$ spectrum, the predictions of our resummation calculation agree well with the CMS data.
The comparision between NLO and resummation calculation is shown in Fig.~\ref{fig:qtF}(b).

\noindent{\bf Summary.} In summary, we have applied the TMD resummation formalism to
study the production of the $Z$ boson  associated with
a high energy  jet at the LHC, where large logarithms of $\ln(Q^2/q_\perp^2)$ were resumed to all orders at the NLL accuracy. 
 We also calculate the NLO total cross section based on the resummation framework and
 the result is slightly different from the MCFM prediction due to the usage of narrow jet approximation in our resummation calculation. 
To ensure the correct NLO total cross section, we have added an additional term 
proportional to  $H^{(0)}$ to account for the above difference in our resummation calculation.  
To compare the prediction of our resummation calculation (for an on-shell $Z$ boson) to the CMS experimental data (with kinematic cuts imposed on $Z$-decay leptons), we approximate the effect of imposing kinematic cuts on the $Z$-decay leptons by employing a reweighting procedure based on the result of PYTHIA8 prediction. 
It shows that we could use a constant reweighting factor to describe the effects of the kinematic cuts imposed on the $Z$-decay leptons.
A detailed  comparison between our resummation calculation and the CMS data is also discussed.
We find that our resummation calculation can describe well the CMS data, both in the distributions of the imbalance transverse momentum ($q_\perp$) and the azimuthal angle ($\phi$) correlation of the $Z$ boson and jet system, for $pp\to Z+jet$ production at the LHC

\noindent{\bf Acknowledgment:}
This work is partially supported by the U.S. Department of Energy,
Office of Science, Office of Nuclear Physics, under contract number
DE-AC02-05CH11231, and by the U.S. National
Science Foundation under Grant No. PHY-1719914. 
C.-P. Yuan is also grateful for the support from the Wu-Ki Tung endowed chair
in particle physics.

\bibliographystyle{apsrev}
\bibliography{reference}

\end{document}